\documentstyle[12pt,epsf]{article}

\begin{document}

\baselineskip=18.8pt plus 0.2pt minus 0.1pt

\makeatletter

\renewcommand{\thefootnote}{\fnsymbol{footnote}}
\newcommand{\beq}{\begin{equation}}
\newcommand{\eeq}{\end{equation}}
\newcommand{\bea}{\begin{eqnarray}}
\newcommand{\eea}{\end{eqnarray}}
\newcommand{\nn}{\nonumber}
\newcommand{\hs}[1]{\hspace{#1}}
\newcommand{\vs}[1]{\vspace{#1}}
\newcommand{\Half}{\frac{1}{2}}
\newcommand{\p}{\partial}
\newcommand{\ol}{\overline}
\newcommand{\wt}[1]{\widetilde{#1}}
\newcommand{\ap}{\alpha'}
\newcommand{\bra}[1]{\left\langle  #1 \right\vert }
\newcommand{\ket}[1]{\left\vert #1 \right\rangle }
\newcommand{\vev}[1]{\left\langle  #1 \right\rangle }

\makeatother

\begin{titlepage}
\title{
\hfill\parbox{4cm}
{\normalsize KUNS-1586\\{\tt hep-th/9907094}}\\
\vspace{1cm}
D0-branes in SO(32)$\times$SO(32) open type 0 string theory
}
\author{Yoji Michishita
\thanks{
{\tt michishi@gauge.scphys.kyoto-u.ac.jp}
}
\\[7pt]
{\it Department of Physics, Kyoto University, Kyoto 606-8502, Japan}
}

\date{\normalsize July, 1999}
\maketitle
\thispagestyle{empty}

\begin{abstract}
\normalsize
We construct D0-branes in SO(32)$\times$SO(32) open type 0 string
theory using the same method as the one used to construct non-BPS
D0-brane in type I string theory. It was conjectured that this theory is
S-dual to bosonic string theory compactified on SO(32)
lattice, which has SO(32)$\times$SO(32) spinor states as excited
states of fundamental string. One of these states seems to correspond
to the D0-brane, and by the requirement that other states which do not
have corresponding states must be removed, we can determine the way of 
truncation of the spectrum. This result supports the conjecture.

\end{abstract}

\vspace{2cm}

PACS codes: 11.25.-w

Keywords: string theory, type 0, D-brane, S-duality, tachyon
condensation, boundary state

\end{titlepage}

Correspondence of SO(32) spinor states is an evidence for the
S-duality between type I and Heterotic SO(32) string
theories. Heterotic SO(32) theory has SO(32) 
spinor states as the first excited states of fundamental string. These 
are the lightest of the states which have SO(32) spinor charge and
therefore cannot decay and must exist in strong coupling regime, which 
is described by type I string theory. Their type I counterpart is non-BPS
D0-brane \cite{s1}.

Let us consider analogous correspondence in type 0 string theory. It was 
proposed that SO(32)$\times$SO(32) open type 0 string
theory \cite{bs,bg} (we will abbreviate it as open type 0 theory) is
S-dual to bosonic string theory compactified on SO(32)
lattice \cite{bg}. In this bosonic
string theory fundamental string has the following worldsheet matter
content:
\beq
 X^\mu(z),\wt{X}^\mu(\bar{z}),\lambda^A(z),\wt{\lambda}^{\wt{A}}(\bar{z}),
 \quad\mu=0,\cdots,9,\quad A,\wt{A}=1,\cdots,32 .
\eeq
Here $A$ is the index of fundamental representation of one SO(32), while
$\wt{A}$ is that of the other SO(32).
The lightest states which have SO(32)$\times$SO(32) spinor charge are
\bea
& \lambda_{-\Half}^A \lambda_{-\Half}^B \lambda_{-\Half}^C \lambda_{-\Half}^D
 \ket{0}_R\ket{\wt{a}}_L,\quad
\lambda_{-\frac{3}{2}}^A\lambda_{-\Half}^B\ket{0}_R\ket{\wt{a}}_L,\quad
\alpha_{-1}^\mu\lambda_{-\Half}^A\lambda_{-\Half}^B\ket{0}_R\ket{\wt{a}}_L, 
& \nn\\
& \alpha_{-1}^\mu\alpha_{-1}^\nu\ket{0}_R\ket{\wt{a}}_L,\quad
\alpha_{-2}^\mu\ket{0}_R\ket{\wt{a}}_L ,& \label{state1}\\
& \wt{\lambda}_{-\Half}^{\wt{A}}\wt{\lambda}_{-\Half}^{\wt{B}}
 \wt{\lambda}_{-\Half}^{\wt{C}}\wt{\lambda}_{-\Half}^{\wt{D}}
 \ket{a}_R\ket{0}_L,\quad
\wt{\lambda}_{-\frac{3}{2}}^{\wt{A}}\wt
 {\lambda}_{-\Half}^{\wt{B}}\ket{a}_R\ket{0}_L,\quad  
\wt{\alpha}_{-1}^\mu\wt{\lambda}_{-\Half}^{\wt{A}}
 \wt{\lambda}_{-\Half}^{\wt{B}}\ket{a}_R\ket{0}_L, 
& \nn\\
& \wt{\alpha}_{-1}^\mu\wt{\alpha}_{-1}^\nu\ket{a}_R\ket{0}_L,\quad
\wt{\alpha}_{-2}^\mu\ket{a}_R\ket{0}_L , & \label{state2}\\
& \ket{a}_R\ket{\wt{a}}_L , & \label{state3}
\eea
where $a$ and $\wt{a}$ are spinor indices of one SO(32)
and the other SO(32) respectively. Here we do not consider the
truncation of spectrum required by modular invariance, etc. We will
return to this point later.

In this paper we construct the type 0 counterpart to these states using the
same method as the one used to construct non-BPS D0-brane in type I
string theory. As we will see, the states corresponding to (\ref{state3}) 
can be found by this method, but the states corresponding to (\ref{state1})
and (\ref{state2}) are not found. This fact suggests what truncation
we should adopt. The result is in accord with the proposal in ref.\
\cite{bg}.

Open type 0 theory is constructed from type 0B theory by $\Omega$
projection, where $\Omega$ is the worldsheet parity inversion, analogously 
to the construction of type I theory from type IIB theory \cite{bg}.
Type 0B theory has two types of RR fields and therefore has two types
of D-branes. We denote their RR charges by $(q,\bar{q})$. Boundary states
of these branes are \cite{bg,bcr}
\beq
\ket{Dp;q,\bar{q}}_0=\frac{1}{\sqrt{2}}\left(\ket{Dp}_{NS+NS+}
 +q\bar{q}\ket{Dp}_{NS-NS-}+q\ket{Dp}_{R+R+}+\bar{q}\ket{Dp}_{R-R-}\right),
\eeq
with
\bea
\ket{Dp}_{NS\pm NS\pm} & = & 
\Half\left(\ket{Dp,+}_{NS}\mp\ket{Dp,-}_{NS}\right), \\
\ket{Dp}_{R\pm R\pm} & = & 
\Half\left(\ket{Dp,+}_{R}\pm\ket{Dp,-}_{R}\right).
\eea
Boundary states of type IIB branes are
\beq
\ket{Dp;q}_{\rm II}=\ket{Dp}_{NS+NS+}+q\ket{Dp}_{R+R+} .
\eeq
For the definition of $\ket{Dp,\pm}_{NS}$ and $\ket{Dp,\pm}_{R}$, and
other notation about boundary states we adopt those of
ref.\ \cite{fgls}. 

Strings stretched between $(q,\bar{q})$ brane and $(\pm q,\pm\bar{q})$ 
brane belong to $\Half(1\pm (-1)^F)$NS sector only, and strings between
$(q,\bar{q})$ brane and $(\pm q,\mp\bar{q})$ brane belong to
$\Half(1-q(-1)^F)$R sector
only \cite{bg,kt}.
Open type 0 theory have 32 D9-branes and 32 anti D9-branes of one type 
for tadpole cancellation. We choose $(1,1)$ and
$(-1,-1)$ as these D9-branes. 
In this theory we can construct two types of D0-branes respectively
from $(1,1)$ D1-brane-$(-1,-1)$ D1-brane system, and $(1,-1)$
D1-brane-$(-1,1)$ D1-brane system, in the same way to construct type I
non-BPS D0-brane from D1-brane-anti D1-brane system (for details see
ref.\ \cite{s1}): 
\begin{enumerate}
\item
Wrap the D1-branes around a compact direction with radius
$R_c=\sqrt{\frac{\ap}{2}}$ and put a ${\bf Z}_2$ Wilson line on one
of the D1-branes.
\item
Define new worldsheet variables
$\phi_R(z), \phi_L(\bar{z}), \phi'_R(z), \phi'_L(\bar{z}), \xi(z),
\wt{\xi}(\bar{z}), \eta(z)$, and $\wt{\eta}(\bar{z})$ :
\bea
& X(z,\bar{z})=X_R(z)+X_L(\bar{z}) , & \\
& \exp(i\sqrt{\frac{2}{\ap}}X_R)=
 \frac{1}{\sqrt{2}}(\xi+i\eta),\quad 
\exp(i\sqrt{\frac{2}{\ap}}X_L)=
 \frac{1}{\sqrt{2}}(\wt{\xi}+i\wt{\eta}) , & \\ 
& \exp(i\sqrt{\frac{2}{\ap}}\phi_R)=
 \frac{1}{\sqrt{2}}(\xi+i\psi),\quad
\exp(i\sqrt{\frac{2}{\ap}}\phi_L)=
 \frac{1}{\sqrt{2}}(\wt{\xi}+i\wt{\psi}) , & \\
& \exp(i\sqrt{\frac{2}{\ap}}\phi'_R)=
 \frac{1}{\sqrt{2}}(\eta+i\psi),\quad
\exp(i\sqrt{\frac{2}{\ap}}\phi'_L)=
 \frac{1}{\sqrt{2}}(\wt{\eta}+i\wt{\psi}) . &
\eea
\item
Give vev to the tachyon field, {\it i.e.} put the Wilson line
$\exp(i\oint dz \frac{1}{2\sqrt{2\ap}}\p\phi \sigma_1)$ along
$\phi$.
\item
Decompactify the compact direction.
$\phi'_D(z,\bar{z})=\phi'_R(z)-\phi'_L(\bar{z})$, $\xi$ and $\wt{\xi}$ are
the variables for this direction with Dirichlet boundary condition.
\end{enumerate}
The only difference between type I D0-brane and type 0 D0-branes in
this construction is that type 0 D0-branes do not have R sector strings.
We can also construct boundary states of type 0 D0-branes following
ref.\ \cite{fgls}:
\begin{enumerate}
\item 
Introduce $\ket{B,\pm}_{NS}$ and $\ket{B,\pm}_R$ for describing D1-brane 
and anti D1-brane with a ${\bf Z}_2$ Wilson line
wrapped around a compact direction with radius $R_c$:
\bea
\ket{B,\pm}_{NS} & = & \ket{D1,\pm}_{NS}+\ket{\bar{D1}',\pm}_{NS} , \\
\ket{B,\pm}_R & = &\ket{D1,\pm}_{R}-\ket{\bar{D1}',\pm}_{R} ,
\eea
where $\bar{D1}'$ means the anti D1-brane with the ${\bf Z}_2$ Wilson line.
\item 
Rewrite these boundary states in terms of the new variables
$\phi(z), \wt{\phi}(\bar{z}), \xi(z),\wt{\xi}(\bar{z}), \eta(z)$, and
$\wt{\eta}(\bar{z})$:
\bea
& X(z,\bar{z})=\Half(X_R(z)+X_L(\bar{z})) ,& \\
& \exp(i\sqrt{\frac{1}{2\ap}}X_R)=
 \frac{1}{\sqrt{2}}(\eta+i\xi),\quad 
\exp(i\sqrt{\frac{1}{2\ap}}X_L)=
 \frac{1}{\sqrt{2}}(\wt{\eta}+i\wt{\xi}) ,& \\ 
& \exp(i\sqrt{\frac{1}{2\ap}}\phi_R)=
 \frac{1}{\sqrt{2}}(\xi+i\psi),\quad
\exp(i\sqrt{\frac{1}{2\ap}}\phi_L)=
 \frac{1}{\sqrt{2}}(\wt{\xi}+i\wt{\psi}) ,&
\eea
\bea
\ket{B,\pm}_{NS/R} & = & \frac{1}{4\pi\ap g_s}
 \sqrt{\frac{2\pi R_c}{\Phi}}
 \exp[-\sum_{n>0}\frac{1}{n}\alpha_{-n}\hat{S}^{(1)}\wt{\alpha}_{-n}]
 \exp[\pm i \sum_{n>0}\psi_{-n}\hat{S}^{(1)}\wt{\psi}_{-n}] \nn\\
 & & \times\exp[-\sum_{n>0}\phi_{-n}\wt{\phi}_{-n}]
 \exp[\pm i \sum_{n>0}\eta_{-n}\wt{\eta}_{-n}]
 \ket{D1,\pm}^{(0)}_{NS/R} \nn\\
 & & \times 2\delta^{(8)}(q^i)\prod_{i=0,2,\cdots,9}^9\ket{k^i=0}
 \sum_{w_\phi={\rm even/odd}}\ket{0,w_\phi} .
\eea
\item 
Put the Wilson line $\exp(i\oint dz
\frac{1}{2\sqrt{2\ap}}\p\phi\sigma_1)$ along $\phi$:
\bea
\ket{B,\pm}_{NS} & \rightarrow &
 \frac{1}{4\pi\ap g_s}\sqrt{\frac{2\pi R_c}{\Phi}}
 \exp[-\sum_{n>0}\frac{1}{n}\alpha_{-n}\hat{S}^{(1)}\wt{\alpha}_{-n}]
 \exp[\pm i \sum_{n>0}\psi_{-n}\hat{S}^{(1)}\wt{\psi}_{-n}] \nn\\
 & & \times\exp[-\sum_{n>0}\phi_{-n}\wt{\phi}_{-n}]
 \exp[\pm i \sum_{n>0}\eta_{-n}\wt{\eta}_{-n}]\ket{0}_{NS} \nn\\
 & & \times 2\delta^{(8)}(q^i)\prod_{i=0,2,\cdots,9}^9\ket{k^i=0}
 \sum_{w_\phi}(-1)^{w_\phi}\ket{0,2w_\phi} ,\\
\ket{B,\pm}_R & \rightarrow & 0 .
\eea
\item 
Rewrite these boundary states in terms of the T-dualized variables
and decompactify the compact direction:
\bea 
\ket{B,\pm}_{NS} & \rightarrow &
 \frac{1}{4\pi\ap g_s}\sqrt{\frac{\pi\ap}{R_c\Phi}}
 \exp[-\sum_{n>0}\frac{1}{n}\alpha_{-n}\hat{S}^{(1)}\wt{\alpha}_{-n}]
 \exp[\pm i \sum_{n>0}\psi_{-n}\hat{S}^{(1)}\wt{\psi}_{-n}] \nn\\
 & & \times\exp[\sum_{n>0}\frac{1}{n}\alpha_{-n}\wt{\alpha}_{-n}]
 \exp[\mp i \sum_{n>0}\psi_{-n}\wt{\psi}_{-n}]\ket{0}_{NS} \nn\\
 & & \times 2\delta^{(8)}(q^i)\prod_{i=0,2,\cdots,9}^9\ket{k^i=0}
 \sum_{w}\ket{0,w} \\
& \rightarrow & \sqrt{2}\ket{D0,\pm}_{NS} .
\label{sqrt2}
\eea
\end{enumerate}
Thus we get two types of boundary states as follows:
\bea
\ket{D1;q,\bar{q}}_0+\ket{\bar{D1}';-q,-\bar{q}}_0 & \rightarrow &
\ket{D0}_{NS+NS+}+q\bar{q}\ket{D0}_{NS-NS-}\equiv\ket{D0;q\bar{q}}_0 .
\label{d0}
\eea
The factor $\sqrt{2}$ in (\ref{sqrt2}) means that the tension of these
D0-branes $T_0$ is $\sqrt{2}$ times the tension of type 0A D0-brane:
\beq
T_0=\sqrt{2}T_0^{{\rm 0A}}=T_0^{{\rm IIA}}=\frac{1}{\sqrt{\ap}g_s} .
\eeq

The rules for computing the spectrum and the interactions of open strings
which end on the D0-branes are the same as in ref.\ \cite{s2} except that
the strings stretched between the same type (different types) of
D0-branes belong to NS (R) sector only.
Similarly, strings between $(\pm1,\pm1)$ D9-branes and $\ket{D0;+1}_0$
of (\ref{d0}) belong to NS sector only, while strings between
$(\pm1,\pm1)$ D9-branes and $\ket{D0;-1}_0$ belong to R sector 
only.
The NS sector gives only massive states because its zero point energy
is $5/8>0$ and the R sector has massless states. 
The R sector massless states belong to SO(32)
fundamental representation corresponding to 32 $(1,1)$ D9-branes or 32
$(-1,-1)$ D9-branes. The zero modes of these massless states form a
Clifford algebra and their quantization gives rise to spinor
representation of SO(32)$\times$SO(32). Therefore $\ket{D0;-1}_0$
corresponds to the state (\ref{state3}). On the other hand, 
$\ket{D0;+1}_0$ has no SO(32)$\times$SO(32) charge and does not
correspond to any state in (\ref{state1}), (\ref{state2}) and
(\ref{state3}).

The type 0 states corresponding to the states (\ref{state1}) and
(\ref{state2}) are not found. It is impossible to construct the states
which have spinor charge of only one SO(32) like the states
(\ref{state1}) and (\ref{state2}) by using boundary states.
This is because the difference between D9 and anti D9-branes is only
the signature of RR part of the boundary states, and it is NSNS part
that can be interpreted by modular transformation as R sector of open
strings which have massless states with SO(32) charge. 

This result suggests what truncation we should adopt. 
What is given in ref.\ \cite{bg} as a ground of S-duality between open
type 0 theory and bosonic string theory on SO(32) lattice is 
the fact that the worldsheet matter content of fundamental string of
bosonic string theory on SO(32) lattice coincides with that of the
counterpart of open type 0 theory. But this leaves two possibilities
in the choice of truncations in bosonic string theory side:
\begin{enumerate}
\item
 We adopt separative GSO projection
 $\Half(1+(-1)^F)\Half(1+(-1)^{\wt{F}})$, where $F$ and $\wt{F}$ are
 the number operators of $\lambda$ and $\wt{\lambda}$ respectively.
\item
 We adopt diagonal GSO projection $\Half(1+(-1)^{F+\wt{F}})$ and in
 addition remove NSR and RNS sectors. This removal is necessary for
 modular invariance of the 1-loop partition function. Indeed the
 partition function is given by
 \bea
  & & \Half(1+(-1)^{F+\wt{F}})({\rm NSNS}+{\rm RR}) \nn\\
  & & = \int \frac{d\tau d\bar{\tau}}{4{\rm Im}\tau} 
  (4\pi^2\ap{\rm Im}\tau)^{-5}\Half
  \frac{|\vartheta_{00}(0,\tau)|^{32}
  +|\vartheta_{01}(0,\tau)|^{32}
  +|\vartheta_{10}(0,\tau)|^{32}}{|\eta(\tau)|^{48}},
 \eea
 which is modular invariant.
\end{enumerate}
In ref.\ \cite{bg} the latter truncation is adopted. 
Diagonal GSO projection leaves the charged tachyon 
$\lambda^A_{-\Half}\wt{\lambda}^{\wt{A}}_{-\Half}\ket{0}_R\ket{0}_L$
which is projected out by separative GSO projection. Condensation of
this tachyon breaks some part of the gauge group. This corresponds to
condensation of the tachyon from string stretched between D9 and anti
D9-branes in open type 0 theory side. 
In addition since the states (\ref{state1}) and
(\ref{state2}) belong to NSR and RNS sector respectively, they are 
removed only by the latter truncation. Therefore we should adopt the
latter truncation. Then the S-duality conjecture in ref.\ \cite{bg} is
supported by the agreement on SO(32)$\times$SO(32) spinor states.

Now we comment on the other branes. Type I string theory has non-BPS
$(-1),7,8$ brane as well as D0-brane \cite{w}. Analogously we can consider
$(-1),7,8$ brane in open type 0 theory. Their boundary states can be
constructed following ref.\ \cite{fgls}:
\bea
\ket{Dp;q}_0 & = &
\frac{\mu_p}{\sqrt{2}}(\ket{Dp}_{NS+NS+}+q\ket{Dp}_{NS-NS-}) ,
\eea
where $\mu_p=2 \;({\rm for}\; p=-1,7),\; \sqrt{2}\; ({\rm for}\; p=8)$.
However, as pointed out in ref.\ \cite{fgls}, because strings
stretched between D9-branes and D7 or D8-brane with $q=1$ have tachyon 
modes , D7 and D8-brane with $q=1$ are unstable. D$(-1)$-branes break
the disconnected components of O(32)$\times$O(32) \cite{w}.

We have considered various branes in the background with tachyons from 
closed strings and open strings stretched between D9 and anti
D9-branes. It is desired to consider these branes in stable background 
without tachyons.

\vs{.5cm}
\noindent
{\large\bf Acknowledgments}\\[.2cm]
I would like to thank S.\ Sugimoto for helpful discussions and M.\ R.\
Gaberdiel for e-mail correspondence.

\newcommand{\J}[4]{{\sl #1} {\bf #2} (#3) #4}
\newcommand{\andJ}[3]{{\bf #1} (#2) #3}
\newcommand{\AP}{Ann.\ Phys.\ (N.Y.)}
\newcommand{\MPL}{Mod.\ Phys.\ Lett.}
\newcommand{\NP}{Nucl.\ Phys.}
\newcommand{\PL}{Phys.\ Lett.}
\newcommand{\PR}{Phys.\ Rev.}
\newcommand{\PRL}{Phys.\ Rev.\ Lett.}
\newcommand{\PTP}{Prog.\ Theor.\ Phys.}
\newcommand{\hepth}[1]{{\tt hep-th/#1}}


\begin{thebibliography}{99}

\bibitem{s1}
 A.\ Sen, ``{\it SO(32) Spinors of Type I and Other Solitons on
 Brane-Antibrane Pair}'',
 \hepth{9808141}, \J{JHEP.}{9809}{1998}{023}

\bibitem{bs}
 M.\ Bianchi and A.\ Sagnotti, ``{\it On the Systematics of Open
 String Theories}'',
 \J{\PL}{B247}{1990}{517} ; 
 A.\ Sagnotti, ``{\it Some Properties of Open String Theories}'',
 \hepth{9509080}

\bibitem{bg}
 O.\ Bergman and M.\ R.\ Gaberdiel, ``{\it A Non-Supersymmetric Open
 String Theory and S-Duality}'',
 \hepth{9701137}, \J{\NP}{B499}{1997}{183}

\bibitem{bcr}
 M.\ Bill\'{o}, B.\ Craps and F.\ Roose, ``{\it On D-branes in Type 0
 String Theory}'',
 \hepth{9902196}, \J{\PL}{B457}{1999}{61}

\bibitem{kt}
 I.\ R.\ Klebanov and A.\ A.\ Tseytlin, ``{\it D-branes and Dual Gauge 
 Theories in Type 0 Strings}'',
 \hepth{9811035}, \J{\NP}{B546}{1999}{155}

\bibitem{fgls}
 M.\ Frau, L.\ Gallot, A.\ Lerda and P.\ Strigazzi, ``{\it Stable
 Non-BPS D-Branes in Type I String Theory}'',
 \hepth{9903123}

\bibitem{s2}
 A.\ Sen, ``{\it Type I D-particle and its Interactions}'',
 \hepth{9809111}, \J{JHEP.}{9810}{1998}{021}

\bibitem{w}
 E.\ Witten, ``{\it D-Branes and K-Theory}'',
 \hepth{9810188}, \J{JHEP.}{9812}{1998}{019}

\end{thebibliography}
\end{document}